\documentclass{svjour2}                    
\smartqed  
\usepackage{graphicx}
\journalname{Gen. Rel. Grav.}
\begin{document}

\title{Large Number Hypothesis}
\subtitle{: A Review}
\titlerunning{Large Number Hypothesis}

\author{Saibal Ray \and Utpal Mukhopadhyay  \and Partha Pratim Ghosh}
\authorrunning{Ray \and Mukhopadhyay \and Ghosh}

\institute{Saibal Ray \at Inter-University Centre for Astronomy
and Astrophysics, Post Bag 4, Pune 411 007, India\\ \emph{Present
address:Physics Department, Barasat Government College, Kolkata
700 124, North 24 Parganas, West Bengal, India}
\email{saibal@iucaa.ernet.in}
  \and
Utpal Mukhopadhyay  \at Satyabharati Vidyapith, North 24 Parganas,
Kolkata 700 126, West Bengal, India
  \and
Partha Pratim Ghosh \at Tara Brahmamoyee Vidyamandir, Matripalli,
Shyamnagar 743 127, North 24 Parganas, West Bengal, India}

\date{Received: date / Accepted: date}

\maketitle

\begin{abstract}
Large dimensionless numbers, arising out of ratios of various
physical constants, intrigued many scientists, especially Dirac.
Relying on the  coincidence of large numbers, Dirac arrived at
the revolutionary hypothesis that the gravitational constant
$G$ should vary inversely as the cosmic time $t$. This hypothesis
of Dirac, known as Large Number Hypothesis (LNH), sparked off many
 speculations, arguments and new ideas in terms of applications.
 Works done by several authors with LNH as their basic platform
 are reviewed in this work. Relationship between some of those works are
 pointed out here. Possibility of time-variations of physical
 constants other than $G$ are also discussed.

\end{abstract}

\section{Introduction}
\label{intro}
Physical constants are not uncommon to us.
During our study of various laws of nature, we come across a
number of constants entangled with those laws. But, it was Dirac
who discovered an apparently unseen thread joining up those
physical constants by a simple yet interesting law, viz., ``Law of
Large Numbers''. Using that law, Dirac arrived at his ``Large
Number Hypothesis'' (LNH) which has profound influence on the
world of physics. In fact, a plethora of works have been done,
both at theoretical and observational level with LNH as their
starting point.

These works can be divided into three broad categories:\\ \quad I.
The first type of works are centered around modification of
Einstein's gravitational theory and related equations for adopting
the idea of $G$ variation as predicted by
LNH~\cite{can77ha,can77aht,can77l,pen04,lau85,rog06,dir73,hoy71n};\\
\quad II. Arguments and counter arguments in favor of and against
LNH for justifying and refuting that hypothesis characterize the
second category of
works~\cite{can78h,can80h,can79ho,can79o301,bis85,bou00,marugan02c,lorenzi03,fal79,nor82,jul83,ste76,ste78,wes81g};\\
\quad III. Testing and applications of LNH can be classified as
the third line of
investigation~\cite{bla78,jor69,egy69,lyt77f,yub82,yub84,van75,dav82,bar79,lod79,gar95him,ber01ta,ber01tb,gil05,ray05m}.

Although so many years have passed after the inception of LNH, it
has not lost its significance. Instead, perhaps it has gained a
new momentum after the discovery of accelerating universe. The
present cosmological picture emerging out of SN Ia
observations~\cite{per98al,rie98al} reveals that the present
universe is accelerating.  Some kind of exotic repulsive force in
the form of vacuum energy is supposed to be responsible for this
acceleration which started about $7$ Gyr. ago. This repelling
force is termed as {\it dark energy} and is designated by
$\Lambda$. On the way of investigating dark energy, many
 variants of $\Lambda$
have been proposed including a constant $\Lambda$. But,
compatibility with other areas of physics demands that $\Lambda$
should be slowly time decreasing. Moreover, the currently
observed~\cite{car02,car02k} small ($\approx 10^{-35}s^{-2}$)
value of $\Lambda$ suggests that it has decreased slowly from a
very high value to its present nearly zero value. This type of
time dependency of $\Lambda$ has a similarity, so far as the main
spirit of the idea is concerned, with that of the gravitational
constant $G$ as proposed by Dirac in his LNH~\cite{pee03r}. Both
$\Lambda$ and $G$ have descended from a very high initial value to
its present small value because the universe is so old. It should
be mentioned here that a long ago, while dealing with large
dimensionless numbers, Eddington~\cite{edd35} proposed a large
number involving the cosmological term $\Lambda$ (then regarded as
a constant geometric term) viz. $ ch(m_nm_e/\Lambda)^{1/2} \simeq
\sqrt N $, where $m_n$ and $m_e$ are the masses of nucleon and
electron respectively, $h$ is the Planck constant, $c$ is the
velocity of light and $N$ is the total number of particles in the
universe. It has also been shown that variable $\Lambda$  models
are generally characterized by particle creation~\cite{car92lw}
which is also a feature of LNH. All these suggest that the link
between LNH and $\Lambda$ can, in no way, be ignored. In fact, in
recent years, a combined framework of LNH and the cosmological
parameter is used to address a number of important issues such as
explanation of flat galactic rotation curves~\cite{whi99k},
unification of LNH with general theory of relativity~\cite{pen05},
possible implications of a variable fine-structure constant
[34,35] etc. Also, the problem of unifying all other forces with
gravity demands an understanding of the coincidence of large
numbers and hence LNH~\cite{pea99}.

On the other hand, for unfolding the true nature of dark energy, a
number of variants of the cosmological term have been suggested,
one of them being kinematical $\Lambda$ of phenomenological
character (for an overview see~\cite{ove98c}). Now, since LNH
predicts that the gravitational constant $G$ should vary with
time, a probable inter-connection between LNH and $\Lambda$ cannot
be ruled out. So an investigation about the behaviour of the scale
factor and other cosmological parameters within the framework of
varying $G$ and $\Lambda$ may be helpful in finding the
significance of LNH in context of the present cosmological
scenario. Such an attempt is made by Ray and
Mukhopadhyay~\cite{ray05m} in their work with phenomenological
$\Lambda$ model under varying $G$. In this review, most of the
works mentioned in this section, will be discussed in subsequent
sections in the relevant contexts. The paper is organized as
follows: idea of large numbers and Dirac's LNH are given in Sec. 2
and 3 respectively. Sec. 4 deals with consequence of LNH in terms
of modifications of gravitational theory and dynamical equations
related to it. Arguments in favor of and against LNH are described
in Sec. 5 while Sec. 6 is devoted to testing and applications of
LNH. Finally, some discussions are being done in Sec. 7.\\

\section{An Overview of the Large Numbers}
\label{sec:1}
 Although Dirac's name is closely associated with large numbers,
 but it was Weyl~\cite{wey17,wey19} who initiated the
 idea of large numbers. He arrived at the number $4 \times 10^{42}$ through a
 comparison between the electron's radius $r_e$ given by ${q_e}^{2}/{4
 \pi \epsilon_0 m_e c^2}$ with the hypothetical radius of a particle
 with a charge $q_e$ and an electrostatic energy equal to that of
 electron's gravitational energy. It is easy to see that the above
 ratio is equal to that of electron's electrostatic Coulomb force
 ($F_e$) and electron's gravitational energy ($F_g$) as well as that of
 the energy of electron's electrostatic field ($E_0$) and electron's
 self-energy ($E_g$) due to its gravitational field. These connections
 prompted Weyl to speculate that the same ratio might also be held
between the radius of the universe to that of an
 electron. Weyl's speculation was found to be nearly correct
 when astronomer Stewart showed~\cite{ste31} that the ratio of
 the radius of the universe and electron was only two
 orders of magnitude smaller ($10^{40}$) than that of Weyl's number.
 In the same year, Eddington~\cite{edd31} speculated that if $N$ be the
 number of particles in the universe, then using the equation
\begin{eqnarray}
\frac{GM_u}{R_u}={C_g}^{2}
\end{eqnarray}

where $M_u$ is the mass of the universe, $R_u$ is the radius of
curvature of the universe, $C_g$ is the maximum speed of recession
of distant galaxies and $G$ is the gravitational constant, one can
write
\begin{eqnarray}
\frac{F_e}{F_g}=\frac{E_0}{E_g}=\frac{q^2}{\pi \epsilon_0 r_e}/\frac{G{m_e}^2}{r_e}=\sqrt N=Weyl's \quad number
\end{eqnarray}
From equation (2), by substituting known values of the parameters,
we can arrive at the number $N = 1.7507 \times 10^{85}$ which is
known as Eddington's magic number. It is to be noted that
Jordan~\cite{jor47} pointed that if $M^*$ and $m_e$ be typical
stellar mass and electron mass respectively, then
\begin{eqnarray}
\frac{M^*}{m_e}\approx 10^{60}\approx (10^{40})^{3/2}.
\end{eqnarray}
        Recently Shemi-zadeh~\cite{she02} has arrived at some large numbers
        of the order of $10^{60}$ by comparing cosmological
        parameters such as Hubble radius ($R_{H}$), mass of the
        universe ($M_{H}$), CMBR temperature ($T_{\gamma}$) with
        their counterparts of the macroscopic world, viz., Planck
        length ($l_P$), Planck mass ($m_P$), Planck energy ($E_P$)
        etc. For instance,
        \begin{eqnarray}
        \frac {R_{H}}{l_P}\approx 10^{60}\approx
        \frac{E_P}{T_\gamma}.
        \end{eqnarray}
        Before leaving this section, special attention should be
        made for three large numbers involving gravitational
        constant $G$, atomic constants $h$, $m_p$ and cosmological
        parameters like Hubble term $H$ and $\rho$, the matter
        density of the observable universe. These three numbers
        are designated by $N_1$, $N_2$, $N_3$ and are given by [55]
        \begin{eqnarray}
        N_1 \equiv \frac{{\bar h}c}{Gm_p^2}\simeq \frac{1}{6}\times 10^{39},
        \end{eqnarray}
         \begin{eqnarray}
        N_2 \equiv \frac{mc^2}{{\bar h}H}\simeq \frac{1}{3}\times 10^{39},
        \end{eqnarray}
         \begin{eqnarray}
        N_3 \equiv \frac{\rho c^3}{m_p H^3}\simeq 10^{79}.
        \end{eqnarray}
        It is easy to see that $N_1$, $N_2$ and $N_3$ can be
        related by
        \begin{eqnarray}
        N_1 \simeq N_2 \simeq \sqrt{N_3}.
        \end{eqnarray}
        Without wasting more time, we conclude this section here
        only mentioning that a list (may not be exhaustive) of
        large numbers is given in Table 1.

\begin{table*}
\begin{minipage}{100mm}
\caption{List of Large Numbers} \label{tab1}
\begin{tabular}{@{}llrrrrlrlr@{}}
\hline Physical Constants Involved  &Large Numbers &References\\

\hline 1. Electrostatic force $F_E$ and gravitational force\\
$F_G$ between a proton and an electron & $F_E/F_G \simeq 10^{40}$
&~\cite{pea99,nar77,and00k}\\ \hline 2. Radius of the universe
$R$\\ and the radius of an electron $r$ &$R/r \simeq 10^{40}$
&~\cite{nar77,cet86ps}\\ \hline 3. Intensity of electromagnetic\\
and gravitational interaction\\ of elementary particles
&$\frac{e^2/{\bar h}c}{G{m_e}^2/{\bar h} c}\simeq 10^{40}$
&~\cite{she02,cet86ps}\\ \hline 4. Mass of a typical star
$M_{\star}$ and electron mass $m_e$&$M_{\star}/m_e \simeq
10^{60}=(10^{40})^{3/2}$ &~\cite{she02}\\ \hline 5. Mass of the
universe $M_U$ and proton mass $m_p$&$M_U/m_p \simeq
10^{80}=(10^{40})^2$ &~\cite{she02}\\ \hline 6. Mass of the
universe $M_U$ and Planck mass $m_P$&$M_U/m_P \simeq
10^{61}$&~\cite{she02}\\ \hline 7. Hubble radius $R_H$ and Planck
length $l_P$&$R_H/l_P \simeq 10^{60}$&~\cite{pea99,she02}\\ \hline
8. Planck mass density $\rho_P$ and the observed\\ matter density
of the universe $\rho$&$\rho_P/\rho \simeq 10^{120} ={N_1}^{1/3}$
&~\cite{she02}\\ \hline 9. Planck energy $E_P$ and CMBR
temperature $T_{\gamma}$ &$(E_P/T_{\gamma})^2 \simeq 10^{60}$
&~\cite{she02}\\ \hline 10. Planck mass $m_P$ and neutrino mass
$m_{\nu}$&$(m_P/m_{\nu})^2 \simeq 10^{61}$ &~\cite{she02}\\ \hline
11. Planck mass $m_P$ and electron mass $m_e$&$(m_P/m_e)^3 \simeq
10^{62}$&~\cite{she02}\\ \hline 12. Planck mass $m_P$ and pion
mass $m_{\pi}$&$(m_P/m_{\pi})^3 \simeq 10^{63}$&~\cite{she02}\\
\hline 13. Planck mass density $\rho_P$ and the current\\ critical
matter density of the universe $\rho_c$&$\rho_P/\rho_c \simeq
10^{121}$ &~\cite{and00k}\\ \hline 14. Electron mass $m_e$ and
Hubble parameter $H$&$m_e c^2/\bar h H \simeq 10^{39}/3$
&~\cite{cet86ps}\\ \hline 15. Matter density of the observed
universe $\rho$,\\ proton mass $m_p$
 and Hubble parameter $H$&$\rho c^3/m_p H^3 \simeq 10^{79}$ &~\cite{cet86ps}\\
 \hline 16. Number of nucleons $N_4$ in the universe&$
\rho_0 (cH^{-1})^3/m_p =10^{80}=(10^{40})^2$ &~\cite{ray79}\\
\hline
 17. Number of proton $N_p$ and baryons $N_b$&
$N_p/N_b \simeq 10^{10} = (10^{40})^{1/4}$ &~\cite{ray79}\\ \hline

\end{tabular}
\end{minipage}
\end{table*}

 \section{Dirac's LNH and its extension}
 Even to a non-initiated person it seems surprising that the
order of the ratios of fundamental constants of the macro and
micro physical world can be so close in so many cases as
demonstrated by equations (1)-(8). So, it is quite natural that a
person like Dirac would be intrigued by those coincidence. Dirac
thought that the coincidence of large numbers was far from being
accidental. He had a firm belief that the coincidence seen among
various cosmological and atomic constants was a manifestation of a
hitherto unknown theory linking up the quantum mechanical origin
of the universe to the various cosmological parameters.
Dirac~\cite{dir37,dir38} pointed out that the ratio of electrical
 ($e^2/4 \pi \epsilon_0 r^2$) and gravitational ($G m_p m_e/r^2$)
 forces between proton and electron in a hydrogen atom is a large
 number of the order of $10^{40}$, i.e.
 \begin{eqnarray}
 \frac{e^2}{4 \pi \epsilon_0 G m_p m_e}\approx 10^{40}
 \end{eqnarray}
 where $e$ is the charge of an electron, $m_p$ is the proton mass
 and $\epsilon_0$ is the permittivity of space.
 Again, the ratio of the then age of the universe ($~2 \times
 10^9$) and the atomic unit of time ($e^2/4 \pi \epsilon_0 m_e c^3$) is
 also nearly of the same size, i.e.
 \begin{eqnarray}
 \frac{4 \pi \epsilon_0 m_e c^3}{e^2}\approx 10^{40}.
 \end{eqnarray}
 Dirac suggested that the two quantities in the left hand side of
 equations (9) and (10) are equal, i.e.
 \begin{eqnarray}
 \frac{e^2}{4 \pi \epsilon_0 G m_p m_e}\simeq \frac{4 \pi \epsilon_0 m_e
 c^3}{e^2}.
  \end{eqnarray}
 Relying on equation (11), Dirac proposed that as a consequence of
 causal connections between macro and micro physical world, some of the
 fundamental constants cannot remain constant for ever; rather
 they should vary with time, however small the change might be.
 This is known as Dirac's LNH. According to LNH, atomic parameters
 cannot change with time and hence the gravitational constant
 should vary inversely with time, expressed in atomic units, i.e.
 \begin{eqnarray}
 G \propto \frac{1}{t}.
 \end{eqnarray}
Also, the hypothesis demands that creation of matter occurs
continuously in the universe. This creation of matter can occur in
two possible ways, viz., ``additive creation'' and
``multiplicative creation''. According to ``additive creation
theory'', matter is created through the entire space and hence in
intergalactic space also. In ``multiplicative creation theory'',
creation of matter occurs only in those places where matter
already exists and this creation proceeds in proportion to the
amount and type of atoms already existing there.
 According to general relativity, $G$ is constant and hence
        we cannot readily consider $G$ as a variable quantity in
        Einstein equation. To overcome this difficulty, Dirac
        considered two metrics. The equations of motion and
        classical mechanics are governed by the Einstein metric
        which remains unaltered while the other metric, known as
        atomic metric, includes atomic quantities and the
        measurement of distances and times by laboratory apparatus~\cite{fau76}.
        The interval $ds(A)$ separating two events as determined
        by apparatus in atomic system of units (a.s.u.) will be different
        from the interval $ds(G)$ between the same two events as
        measured in the gravitational system of units (g.s.u.).
 This implies that equations written in g.s.u. and a.s.u. cannot be
        used at a time until one of them is converted to the other
        system of units~\cite{rog06}.
          The velocity of light is unity for both metrics.
        Considering the case of a planet orbiting the sun, Dirac~\cite{dir74}
        showed that the relationship of Einstein and atomic
        metric was different for additive and multiplicative
        creation theory. In terms of the atomic distance scale,
        the solar system is contracting for the additive creation model
        while it is expanding in multiplicative creation.

Zeldovich~\cite{zel68} extended Dirac's LNH by including the
cosmological parameter in its realm and defined $\Lambda$ by
 \begin{eqnarray}
|\Lambda| = \frac{8\pi G^2 m_p^6}{h^4}.
 \end{eqnarray}
Then Zeldovich showed that $\Lambda$ produces the same
gravitational field in the vacuum as that produced by matter in
space and hence the cosmological term should be included in the
field equations as full fledged term in the presence of ordinary
matter. The gravitational energy of the vacuum was interpreted by
Zeldovich as interactions of virtual particles separated by a
distance $h/m_pc$ and the amount of energy created by the
gravitational interactions of these particles is given by
 \begin{eqnarray}
\epsilon = \frac{Gm_p^6 c^4}{h^4}.
 \end{eqnarray}
LNH was further extended by Sakharov~\cite{sha67} who proposed a
gravitational theory based on the consideration of vacuum
fluctuations. He suggested that there should be a fundamental
length $\sim 10^{-33} cm^{-1}$, less than which the theory is not
valid. According to Matthews~\cite{mat98}, in the extended LNH,
the mass of a proton (chosen by Dirac) should be replaced by the
effective mass of the vacuum energy density of the relevant epoch
because this replacement would agree with experimental results. It
has already been mentioned that Eddington~\cite{edd35} linked up
the cosmological parameter $\Lambda$ with the total number of
particles of the universe through the relation
 \begin{eqnarray}
ch(m_n m_e/\Lambda)^{1/2} \simeq \sqrt N.
\end{eqnarray}
Above relation tells us  readily that if $N$ increases with the
age of the universe, then $\Lambda$ is also time dependent.
Berman~\cite{ber94} has called this whole idea as Generalized LNH.

 \section{Consequences of Dirac's LNH}

  \subsection{Variable $G$ Cosmology}
   Dirac's LNH has many significant consequences and as such
   depending on it, a plethora of works have been done. Most of
   these works are centered around the variation of $G$.

    \subsubsection{Scale-covariant Theory}
     It is well known that Einstein's equations of general
   relativity do not permit any variation in the gravitational
   constant G because of the fact that the Einstein tensor has zero divergence and
   by energy conservation law $T_{; \nu}^{\mu\nu}$ is also zero.
   So, in the light of Dirac's LNH, some modifications of Einstein
   equations are necessary. This is because, if we simply allow $G$ to be a
   variable in Einstein equations, then energy conservation law is
   violated~\cite{can77aht,wes81g}. So, the study of the effect of varying
   $G$
   can be done only through modified field equations and modified
   conservation laws. For this purpose, Canuto et al.~\cite{can77ha,can77aht}
   developed a scale-covariant (also termed as scale-invariant)
   theory. In this theory a gauge function $\beta$ is chosen and
   the essence of scale-covariant theory lies in the fact that
   physical laws remain unaffected by the choice of the gauge
   function. Scale-covariant theory is developed by scale
   transformation and using different dynamical systems for
   measuring space-time distances. Let us start with the
   conventional Einstein equation
  \begin{eqnarray}
   \bar{G}_{\mu\nu} = -8\pi \bar{T}_{\mu\nu}+ \Lambda \bar{g}_{\mu\nu}
  \end{eqnarray}
   having the line element
  \begin{eqnarray}
   {\bar{ds}}^2 = {\bar g}_{\mu\nu} dx^{\mu} dx^{\nu}.
   \end{eqnarray}
   Using the transformation
  \begin{eqnarray}
   ds = \beta^{-1}(x)\bar {ds}
   \end{eqnarray}
   and by considering necessary modification of Ricci
   tensors~\cite{eis26} (i.e. of Einstein tensors as well) it has been shown
   [2] that the field equations, in general units, can be written
   in modified form as
 \begin{eqnarray}
  G_{\mu\nu}+ 2\frac{\beta_{\mu ; \nu}}{\beta}-
  4\frac{\beta_\mu \beta_\nu}{\beta^2}-
  g_{\mu\nu}\left(2\frac{\beta^{\lambda}_{;\lambda}}{\beta}-
  \frac{\beta^{\lambda} \beta_{\lambda}}{\beta^2}\right) = -8\pi
  T_{\mu\nu}+ \Lambda g_{\mu\nu}
  \end{eqnarray}
  where $T_{\mu\nu}$ is the energy-momentum tensor and $\Lambda$
  is the so called cosmological term which is related to $\bar{\Lambda}$ by the
  relation
 \begin{eqnarray}
  \Lambda = \beta^2 \bar{\Lambda}.
   \end{eqnarray}
  It is also shown that energy conservation law can be written in
  a modified form as
 \begin{eqnarray}
  \dot{\rho}+ (\rho+p)u^{\mu}_{; \mu} =-\rho \left(\frac{\dot G}{G}+
  \frac{\dot{\beta}}{\beta}\right)- 3p\frac{\dot{\beta}}{\beta}.
  \end{eqnarray}
  The corresponding Friedmann equations with the conventional
  Robertson-Walker metric
\begin{eqnarray}
 ds^2 = dt^2- R^2(t)\left(\frac{dr^2}{1-kr^2}+ r^2 d\theta^2+ r^2 sin^2\
\theta d\phi^2\right)
 \end{eqnarray}
 reduces to the form
\begin{eqnarray}
\left(\frac{\dot R}{R}+ \frac{\dot{\beta}}{\beta}\right)^2+
\frac{k}{r^2} = \frac{8\pi G\rho}{3}+ \frac{\Lambda}{3},
\end{eqnarray}
\begin{eqnarray}
\frac{\ddot R}{R}+ \frac{\ddot{\beta}}{\beta}+
\frac{\dot{\beta}}{\beta}\frac{\dot R}{R}-
\frac{\dot{\beta}^2}{\beta^2} = -\frac{4\pi G}{3}(3p+\rho)+
\frac{\Lambda}{3}.
\end{eqnarray}
The energy conservation laws, in this case, become
\begin{eqnarray}
\dot{\rho}+ 3\frac{\dot R}{R}(\rho+p) = -\frac{\rho}{G \beta}
\frac{d}{dt}(G \beta)- 3p \frac{\dot{\beta}}{\beta}.
\end{eqnarray}

 However, in the scale-covariant theory~\cite{can77ha,can77aht}, the cosmological term
$\Lambda$ is not a constant rather it varies like $\beta^2$. Now,
determination of $\beta(t)$ is very crucial for scale-covariant
theory because determination of $\beta(t)$ enables us to compare
that theory with observational results. As $\beta(t)$ gives us the
liberty in choosing the system of units, it is not possible to
determine it within the theory. This means that we cannot
formulate any dynamical equation for determining $\beta(t)$ and
hence imposition of external constraint is obligatory here.
Relation with gauge fields and the cosmological parameter suggests
that $\beta$ is inversely proportional to t. Again, adopting LNH,
it can be shown that $\beta$ can alternatively be written as
$\beta = {t_0}/t$, where $t_0$ is the present age of the
universe~\cite{can77ha,can77aht,can77l}.\\
              Using their scale-covariant cosmology,
Canuto and Hsieh~\cite{can78h} have shown that it is possible to
reconcile $3K$ blackbody radiation with Dirac's LNH and scale
factor $R(t)$ as well as curvature constant $k$ can be determined
without using $m$ versus $z$ relation or any other classical
cosmological test. It has also been demonstrated
there~\cite{can78h} that $k=0$ i.e. universe is flat and the
relation between $G$ and the gauge function $\beta(t)$ is given by
\begin{eqnarray}
G \beta^2 = constant.
\end{eqnarray}
It may be mentioned here that recent observational results
provided by WMAP, COBE etc. support scale-covariant cosmology
regarding the geometry of the universe.

Being intrigued by Dirac's LNH Peng~\cite{pen05} has modified
Einstein's general relativity theory by allowing $G$ to be a
variable (as suggested by Dirac) so that the new theory can be
applied to cosmology without any inconsistency. In Peng's work, a
tensor term arising from variation of $G$ plays the role of the
cosmological term $\Lambda$. Moreover, natural constants which
evolve with time (viz., $m_e, m_p, e,\bar h, K_B$ etc.) are
modified there~\cite{pen05} systematically so that the new
constants remain really constant for sufficiently long time. In
another work, Peng~\cite{pen04} has modified Einstein's general
theory of relativity by considering the gravitational constant $G$
as a variable. This modification is achieved by including a tensor
term which crops up naturally from the derivative of $G$ and not
from the cosmological term $\Lambda$. Unlike Dirac, Peng assumed
that $m_e, m_p$ and $e$ are not constants but evolve according to
the rules $Gm\propto t$ and $e^2\propto tm_e$ where $m$ may be any
one of $m_e, m_p$ or $M$, the total mass of the universe. Then, by
making a fundamental assumption of the form
\begin{eqnarray}
\phi^2 = (G/G_0)^{1/n} = (t/t_0)^{-1}
\end{eqnarray}
where $\phi^2$ is a dimensionless variable, Peng~\cite{pen04}
showed that in the modified Einstein equation the cosmological
term $\Lambda_{\alpha}^{\beta}$ is determined by $\phi$. For the
usual FLRW metric, the modified field equations become
\begin{eqnarray}
R^3\frac{d\phi^2}{dt} = (3-\omega/2)^{-1}8\pi
G_0\tilde{\rho_m}(t')R^3(t')t,
\end{eqnarray}
\begin{eqnarray}
\phi^2\frac{-3(\dot R^2+k)}{R^2}-3\phi^2\dot
R/R-\omega\dot\phi^2/2 = \frac{-8\pi G_0
\tilde{\rho}(t')R^3(t')}{R^3}
\end{eqnarray}
where $t'$ is an arbitrary constant argument.

It has been shown by Peng~\cite{pen04} that this modified theory
is consistent with LNH. Moreover, Hubble's relation, derived from
this new theory is compatible with observational results.
Einstein's theory is shown to be a special case of the modified
theory for phenomena of short duration extended to short
distances.

 For reconciling LNH with
Einstein's theory of gravitation, Lau~\cite{lau85} selected the
general form of Einstein's field equations, viz.,
\begin{eqnarray}
R^{\mu\nu}- (1/2)g^{\mu\nu}R+ \Lambda g^{\mu\nu}= -8\pi
GT^{\mu\nu}
\end{eqnarray}
and showed that for a time-dependent $G$ and $\Lambda$, LNH
satisfies equation (30). According to Lau~\cite{lau85}, time
varying $G$ is important only for the early period or for knowing
the entire evolutionary history of the universe. But at a time
sufficiently away from the Big-Bang, $G$ can conveniently be
regarded as a constant. He argues also that at the present epoch,
cosmological term $\Lambda$ has a very small value and hence can
be approximated to zero. With a constant $G$ and zero $\Lambda$,
equation (30) reduces to the conventional form
\begin{eqnarray}
R^{\mu\nu}-(1/2)g^{\mu\nu}R= -8\pi GT^{\mu\nu}
\end{eqnarray}
of Einstein's field equations. It is to be noted that being
motivated by LNH and assuming time dependent $\Lambda$ and $G$,
Lau and Prokhovnik~\cite{lau86p} have developed a scalar-tensor
theory in terms of an action principle as a modification of
Einstein's theory.

\subsubsection{Dirac's theory of variable G}
 Another G-variable theory is
Dirac's theory~\cite{dir37,dir73} which is a theory of the
background structure of the universe and its geometrical
properties are described by a gauge function $\beta$ which
determines the ratio of gravitational units and electromagnetic
units~\cite{wes81g}. Dirac's theory is related to the question of
a zero or non-zero cosmological term $\Lambda$. Field equations
and LNH imply that two natural gauges are associated with Dirac's
theory~\cite{dir37,dir73}. One gauge, where $\Lambda = 0$, is
known as additive creation (zero gauge) theory. Another gauge in
which $\Lambda$ is finite is known as multiplicative creation
theory. So, continuous creation is an integral part of Dirac's
theory.

\subsubsection{Hoyle-Narlikar theory}
Hoyle-Narlikar theory~\cite{hoy71n} is based on two main
principles, viz., conformal invariance and the absorber theory of
radiation~\cite{wes81g}. For conformal invariance, natural laws
remain invariant under changes of the gauge function $\beta$ and
hence possesses a technical similarity with Dirac's theory. The
second principle is a theory of electromagnetism and has been
shown to be an integral part of cosmology by Hoyle-Narlikar
theory.

\section{Status of Dirac's LNH}
\subsection{Arguments against Dirac's LNH} Time and again, a number
of objections have been raised against Dirac's
LNH~\cite{dir37,dir38}. According to Falik~\cite{fal79}, $3K$
blackbody radiation cannot be reconciled with LNH, not even in the
version of LNH that is constructed by taking into account the
blackbody radiation. Because, LNH cannot explain the large amount
of helium present in the universe. Again, a prediction of Dirac's
LNH is that the number of nucleons in the universe should vary as
$t^2$ where $t$ is the cosmic age. Norman~\cite{nor82}, citing
some examples, has shown that multiplicative creation theory is in
huge disagreement with observational results regarding isotropic
abundance ratios of some elements. For instance, it has been
found~\cite{boa54,vdo71m} that $13_C/12_C$ is $1.1\times 10^{-2}$
while according to LNH~\cite{nor82} it is as large as $4.3$. Also,
according to Norman's~\cite{nor82} calculation $40_K/39_K$ ratio
obtained from Dirac's LNH is $14$ while in reality, it is
$1.3\times 10^{-14}$~\cite{bur66lw}. Thus, in both the cases the
values as provided by LNH are much higher than the actual values.
Another point of objection against Dirac's LNH is self-consistent
within the universe of Einstein de-Sitter model. If only curvature
constant and the cosmological term are both zero, then Dirac's
theory is compatible with the well known Friedmann-Lema\^{i}tre
model. Otherwise, the relation $N_1 = N_2 \simeq \sqrt N_3$ should
be considered as accidental and hence no inference can be drawn
regarding variation of $G$ and continuous creation of
matter~\cite{jul83}. Many cosmological models proposed by Dirac
require photon creation or destruction. On the other hand, for any
cosmological model that obeying Cosmological Principle and
evolving in time, number of photons in a co-moving volume must be
conserved. Steigman~\cite{ste78} has shown that cosmological
models based on Dirac's LNH are in sharp disagreement with the
standard Planck spectrum. For verifying the plausibility of
continuous creation of matter as predicted by Dirac relying on his
LNH, Steigman~\cite{ste76} developed cosmological models with two
modes of particle creation. It has been shown there that creation
of particles is unnecessary because, as it is expected, the number
of particles $N$ varies as $t^2$ where $t$ is the cosmic age.

\subsection{Arguments in favor of Dirac's LNH}
However, many workers have supported Dirac's LNH by raising
various counter-arguments. Canuto and Hsieh~\cite{can78h} have
provided a physical basis needed for Dirac's cosmology and have
derived the relevant dynamical equations. It has also been shown
by them~\cite{can78h} that LNH is not in disagreement with $3K$
blackbody radiation, rather it is of utmost importance for
predicting the scale factor and the curvature constant. In another
work, Canuto and Hsieh~\cite{can80h} have refuted
Falik's~\cite{fal79} argument by showing that Falik's conclusion
is based on two faulty assumptions, viz.,
 \begin{eqnarray}
 \rho_\gamma \sim T^4,
\end{eqnarray}
 \begin{eqnarray}
 RT = constant
\end{eqnarray}
where $\rho_\gamma$ is the energy density of radiation in local
thermodynamic equilibrium, $T$ is the equilibrium temperature and
$R$ is the scale factor of RW metric. According to Canuto and
Hsieh~\cite{can80h}, correct relations should have been
\begin{eqnarray}
\rho_\gamma \sim \beta^2 G^{-1} t^4,
\end{eqnarray}
\begin{eqnarray}
\beta RT = constant.
\end{eqnarray}
Moreover, in the opinion of Canuto and Hsieh~\cite{can80h} the
relation $G \sim t^{-1}$ should not be extrapolated backwards in
the early epoch as done by Falik~\cite{fal79} because, this
relation is compatible with observational results only in the
present matter dominated universe~\cite{can79ho,can79o301}.
Bishop~\cite{bis85} has opposed Julg's~\cite{jul83} argument by
stating that one should not assume LNH and $G\rho R^3 \sim
constant$ simultaneously because LNH cannot be discussed within
FRW models. Not only that, the relation $G\rho R^3 \sim constant$
is inconsistent with two other relations proposed by
Dirac~\cite{dir38,dir37,dir74}, viz.,
\begin{eqnarray}
N \sim \rho R^3 = constant,
\end{eqnarray}
\begin{eqnarray}
GN \sim t
\end{eqnarray}
where $N$ is the number of nucleons in the universe.

LNH can be used to determine the gauge function $\beta(x)$
involved in the scale covariant theory  developed by Canuto and
others~\cite{can77ha,can77aht}. It was previously found that this
$\beta(x)$, when applied in computations of stellar evolution,
support multiplicative creation but not additive creation.
Slothers~\cite{slo76} made a rough calculation for determining the
effects of multiplicative creation on the luminosity of a white
dwarf and showed that the lower bound of the luminosity was at
least one order of magnitude higher than that which was observed
physically. But, through a detailed calculation,
Lodenquai~\cite{lod79} has refuted Slothers's conclusion regarding
lower bound of white dwarf luminosity and hence viability of
multiplicative creation theory is not lost. Relying on his
expansion centre model~\cite{lor00a,lor00b} related to Milky Way,
Lorenzi~\cite{lorenzi03} has been able to develop a cosmological
picture of the universe which is shown to agree with Dirac's LNH.
Chao-wen and Slothers~\cite{cha75s} have shown that Dirac's
multiplicative creation theory do not violate any observed fact
about sun, but theory of additive creation is not of that status.
Genreith~\cite{gen99} has shown that Dirac's LNH can be explained
by a fractal model of the universe. Moreover, Dirac's conjecture
regarding an inter-connection between micro and macro universe and
that of Einstein about the role of gravitational fields in the
structure of elementary particles can be reconciled using that
model~\cite{gen99}.

However, Holographic principles can be helpful for explaining
Dirac's LNH. In very simplified form, holographic principle states
that the entropy $S$ (actually, $S$ is the entropy divided by
$\sigma$, the Boltzmann constant) of a physical system subject to
gravity is bounded from above by a quarter of its boundary area in
Planck units i.e.
\begin{eqnarray}
S \leq \frac{A}{4 l_p^2}.
\end{eqnarray}
Also, $\bar h$, $G$ and $c$ provide a natural system of units of
length and mass given by
\begin{eqnarray}
l_p = \sqrt \frac{{\bar h} G}{c^3},
\end{eqnarray}
\begin{eqnarray}
m_p = \sqrt \frac{{\bar h}c}{G}.
\end{eqnarray}
According to Bousso~\cite{bou00} the holographic principle leads
to the prediction that the number of degrees of freedom $N$
available in the universe is related to the cosmological parameter
$\Lambda$ by the relation
\begin{eqnarray}
N = \frac{3\pi}{\Lambda l_p^2 ln 2}.
\end{eqnarray}
The $N$ bound conjecture states that, $S$ is bounded by $N ln 2$.
Marugan and Carneiro~\cite{marugan02c} have shown that if one
assumes a homogeneous, isotropic and flat universe dominated by
the cosmological term $\Lambda$, then Dirac's large number
coincidence can be explained in terms of the holographic $N$ bound
conjecture.

Another objection raised against $G \propto {1/t}$ relation comes
from stellar astrophysics. The problem is like this: since stellar
luminosity is proportional to $G^7 M_{\star}^5$~\cite{wes81g},
then variation of $G$ implies an abnormally high solar luminosity
which does not fit with observations. But, according to Wesson and
Goodson~\cite{wes81g} the relation $GM_{\star} =
constant$~\cite{can77aht} leads to the result that $L$ is nearly
constant. If this consideration, along with the change of
earth-sun distance due to change in $G$ (and hence possible change
in mass as well), is taken into account then this problem of
variable $G$ cosmology may be solved.

Davidson~\cite{dav82}, in an attempt for testing Dirac cosmology
in the light of observational results, has shown that if one
assumes Dirac's LNH and admits the existence of two metric scales,
viz., atomic scale or $A$ scale and Einstein scale or $E$ scale
then many important observational cosmological features like
Hubble's parameter, the cosmic age, the cosmic mass density follow
naturally. It is to be noted here that Dirac's LNH-based cosmology
differs from that of canonical general relativistic ({\it GR})
cosmology. For instance, according to Dirac cosmology, the atomic
age of the universe is $T/3$ where $T$ is the observed Hubble
time, whereas in General Relativity it is $2T/3$ for $\Omega_0 =
1$.

\section{Application and Testing of LNH}

\subsection{Cosmology and Astrophysics}
Following Dirac's~\cite{dir37,dir38} approach regarding large
number coincidence, Barrow~\cite{bar79} has predicted proton
half-life period. His prediction can be compared with experimental
results for verification. Relying on the experimental results of
the time variation of the fine-structure
constant~\cite{web99al,web01al}, Berman and
Trevisan~\cite{ber01ta} considered a model in which electric
permittivity $\epsilon_0$ and magnetic permeability $\mu_0$ vary
with time such that the speed of light remains constant. Using
LNH, it has been possible to judge the time dependency of $N$, the
number of nucleons in the universe~\cite{ber01ta}. Moreover, value
of the deceleration parameter as estimated from the same
investigation is compatible with the supernova
data~\cite{per98al,rie98al} while the calculated value of the
cosmological term falls within the acceptable range. In another
work, relying on the experimental data of Webb et
al.~\cite{web01al} regarding the time variation of the fine
structure constant, Berman and Trevisan~\cite{ber01tb} have
derived possible time variations of some other parameters of the
universe, viz., number of nucleons in the universe, the speed of
light, gravitational constant and the energy density. It has been
possible to calculate~\cite{ber01tb} the value of the deceleration
parameter which points towards an accelerating universe and hence
supports SN Ia data~\cite{per98al,rie98al}.

It should be mentioned here that Gomide~\cite{gom76} studied
cosmological models with varying $c$ and (or) varying $\epsilon_0$
including LNH in his framework of study. Berman and
Trevisan~\cite{ber01ta} elaborated a full model containing a
Jordan-Brans-Dicke (JBD) framework with time-varying speed of
light. In another work, Berman and Trevisan~\cite{ber01tb} have
commented that similar conclusions could be attained by applying
Dirac's LNH with $c=c(t)$.

Being intrigued by the suggestion of Dirac, regarding time
variation of $G$, Garcia-Berrow et al.~\cite{gar95him} have
investigated about the effect of a decreasing $G$ on the cooling
rate of white dwarfs. It has been shown there that variable $G$
strongly affects the cooling rate of white dwarfs at low
luminosity. The star expands and cooling process gets accelerated
due to decrease in $G$. For two separate cases, two upper bounds
for $\dot G/G$ are derived~\cite{gar95him} which are in good
agreement with those obtained from binary pulsar.

Using the Weak Field Approximation, Whitehouse and
Kraniotis~\cite{whi99k} have determined the value of the
cosmological parameter $\Lambda$ from galactic rotation curves and
have found that it agrees with their theoretically derived value
of the same parameter using extended LNH. In the same paper,
values of other cosmological parameters, viz., gravitational
constant $G$, gravitational modification constant and the
effective mass density were predicted. They have also shown that
within the extended LNH, only two parameters, viz., fundamental
length and the vacuum energy density are sufficient to completely
specify the cosmological parameters for that epoch. This approach,
according to them~\cite{whi99k}, may be helpful for finding a
fundamental theory linking up the atomic and cosmological
parameters and Dirac's dream may come true. Moreover, it has been
shown by them~\cite{whi99k} that the flat rotation curve of
galaxies may be explained by the cosmological term $\Lambda$ and
presence of dark matter is not necessary if Newton's gravitational
equation is modified in the form
\begin{eqnarray}
F_m = -\frac{Gmm_0}{r^2}+ G_{\Lambda} mr
\end{eqnarray}
where $G_{\Lambda}$ is the gravitational force exerted by the
cosmological term $\Lambda$ and represents a fifth fundamental
force which is directly proportional to the distance. However, the
value of $\Lambda$ derived in~\cite{whi99k} is negative and hence
represents a decelerating universe, which goes against the modern
picture of an accelerating universe with a repulsive cosmological
term.

Gilson~\cite{gil05} has shown that implications of Dirac's LNH can
be directly derived from his quantum theory of
gravitation~\cite{gil04} and three large numbers $\alpha\bar
h/Gm_pm_e$, $c/Hr_e$ and $N$ can be expressed as three closed
formulae with definite coefficients involving other known physical
constants. Secondly, from the theory mentioned above, it has been
possible to develop two quantum Friedmann cosmologies in which the
cosmological parameter $\Lambda$ plays a very basic and
fundamental role having nice agreement with measurement. Dahnen
and Honl~\cite{deh69h} has shown that variability of $G$, as
proposed by Dirac, suggests that QSO's are normal galaxies at
their early stages and vice-versa.

 Constructing two solar models for testing two
types of matter creation of Dirac's cosmology, Carignan et
al.~\cite{car79bs} have shown that the first model fits very well
with the theory of multiplicative creation of Dirac while the
second model does not support the modified multiplication
theory~\cite{car78} in the sense that it shows an excess solar
neutrino flux and presence of excessive hydrogen on the surface.
Assuming Dirac's multiplicative theory (i.e. $G\propto t^{-1}$ and
$M\propto t^2$) VandenBerg~\cite{van77} has theoretically
calculated isochrones and luminosity functions for old stellar
systems. His results do not show any difference from normal
stellar evolution regarding colour-magnitude diagram. Applying his
scalar-tensor theory to a cosmological model which obeys LNH, Lau
and Prokhovnik~\cite{lau86p} have deduced the time dependent form
of the cosmological parameter. A viable explanation for the
smallness of $\Lambda$ is also provided by them along with the
possible significance of the scalar field~\cite{lau86p}.

\subsection{Impact on Planetary Science}
 LNH has cast its shadow in the field of planetary science also.
 According to the literature, LNH has tremendous influence on
 the rotation of the earth. Blake~\cite{bla78} has shown
that early Dirac, additive creation and multiplicative creation
theories have different predictions about variations of lengths of
month and year. Comparing fossil data regarding these variations
with those predicted by LNH, Blake~\cite{bla78} has suggested that
early Dirac and additive creation versions of LNH cannot be
accepted here while multiplicative creation theory fits well with
observations.

LNH and hence variation of $G$ has played a role in the
calculation of earth's rate of expansion also. The average rate of
expansion, calculated by several geophysicists, is $0.48$ mm. per
year. Such a large expansion cannot be explained in terms of
geophysical processes. Jordan~\cite{jor69} and Egyed~\cite{egy69}
 proposed that the amount of expansion may be explained if one
admits the variability of $G$, as suggested by Dirac. Lyttleton
and Fitch~\cite{lyt77f}, taking into consideration a variable $G$,
calculated the change of radius of earth consisted of a liquid
core and a mantle. Yubushita~\cite{yub82} constructed an earth
model with a liquid core, a mantle and an outer shell and obtained
the following relationships between rates of change of earth's
radius and that of $G$ for three versions of Dirac's LNH:

(i) For no creation model,
 \begin{eqnarray}
\frac{\dot R}{R} = -0.062 \times \frac{\dot G}{G},
\end{eqnarray}
(ii) For additive creation,
 \begin{eqnarray}
\frac{\dot R}{R} = -0.33 \times \frac{\dot G}{G},
\end{eqnarray}
(iii) For multiplicative creation,
 \begin{eqnarray}
\frac{\dot R}{R} = -0.61 \times \frac{\dot G}{G}.
\end{eqnarray}
Taking $\dot G/G = -1/t$ and using the value of the Hubble
parameter as $6 \times 10^{-11} yr^{-1}$, he finally obtained
$\dot R = 7 \times 10^{-3} cm~yr^{-1}$ for no creation, $\dot R
1.9 \times 10^{-2} cm~yr^{-1}$ for additive creation and $\dot R
2.5 \times 10^{-2} cm~yr^{-1}$ for multiplicative creation models.
According to Yubushita~\cite{yub82}, both no creation and additive
creation models are inadequate for explaining the calculated rate
of earth's expansion. The third option, i.e. multiplicative
creation theory can be consistent with the observational data if
the value of the Hubble parameter $H$ be taken to be $10^{-10}
yr^{-1}$. In another paper, Yubushita~\cite{yub84} has also shown
that if $G$ changes in accordance with Dirac's LNH, then the
radius of the primordial earth would have been $700$ km. less than
the present value. Of course, under some special assumptions, this
change in radius is shown to be energetically feasible. Analyzing
lunar occultation data since 1955, Van Flandern~\cite{van75}
calculated moon's mean longitude. Faulkner~\cite{fau76} showed
that Van Flandern's result is consistent with Dirac's additive
creation theory.

\subsection{Amount of Variation of G}
 Ever since Dirac's proposition of a possible
time variation of $G$, a volume of works has been centered around
the act of calculating the amount of variation of the
gravitational constant. Gaztanaga et al.~\cite{gaz02al}, relying
on data provided by SN Ia~\cite{per98al,rie98al} have shown that
the best upper bound of the variation of G at cosmological ranges
is given by
 \begin{eqnarray}
-10^{-11} \leq \left|\frac{{\dot G}}{G}\right| \leq 0
\end{eqnarray}
where $z$, the red-shift, assumes the value nearly equal to $0.5$.
Observation of spinning-down rate of pulsar PSR J2019+2425
provides the result~\cite{arz95,sta03}
\begin{eqnarray}
\left|\frac{{\dot G}}{G}\right| \leq {(1.4-3.2)} \times 10^{-11}
yr^{-1}.
\end{eqnarray}
Depending on the observations of pulsating white dwarf star $G$
117-B 15A, Benvenuto et al.~\cite{ben04al} have set up the
astereoseismological bound on $\dot G/G$ as
 \begin{eqnarray}
- 2.50 \times 10^{-10} \leq \left|\frac{{\dot G}}{G}\right| \leq 4
\times 10^{-10} yr^{-1}
\end{eqnarray}
while using the same star Biesiada and Malec~\cite{bie04m} have
shown that
 \begin{eqnarray}
\left|\frac{{\dot G}}{G}\right| \leq 4.1 \times 10^{-11} yr^{-1}.
\end{eqnarray}
Using Nordtvedt's~\cite{nor70} expression for $\dot G/G$ in
generalized Brans-Dicke theory, Sahoo and Singh~\cite{sah03s} have
shown that for some particular values of the coupling constant
$\omega_0$, viz., $-1.9$ or $-1$, the numerical value of $\dot
G/G$ at present is about $2 \times 10^{-10}$ per year which lies
within the observational limit. Various ranges of time variations
of $G$, provided by both theoretical and observational results are
enlisted in Table 2.

\begin{table*}
\begin{minipage}{100mm}
\caption{Values of $\alpha$ and $\beta$ for average $\dot G/G$
when
  $t_0=14$ Gyr, $\Omega_m=0.3$, $\Omega_{\Lambda}=0.7$ and $z \simeq 0.5$}
\label{tab2}
\begin{tabular}{@{}llrrrrlrlr@{}}
\hline Ranges of $\dot G/G$ yr$^{-1}$         &Sources &$\alpha$
&$\beta$\\ \hline $-(1.10 \pm 1.07) \times 10^{-11}<\frac{\dot
G}{G}<0$&PSR $1913 + 16$~\cite{dam88al} &-0.0852&0.4074\\ \hline
$-1.60 \times 10^{-12}<\frac{\dot G}{G}<0$&Helioseismological
data~\cite{gue98al}& -0.0115 &0.0670\\ \hline $(-1.30 \pm 2.70)
\times 10^{-11}$&PSR
B1855+09~\cite{arz95,kas94tr}&-0.1023&0.4698\\ \hline $-1.40
\times 10^{-11}<\frac{\dot G}{G}<+2.60 \times
10^{-11}$&Loren-Aquilar et al.~\cite{lor03al}&0.0410 &-0.2812\\
\hline $-10^{-11} \leq \frac{\dot G}{G}<0$&Supernove Type
Ia~\cite{gaz02al} &-0.0769&0.3750\\ \hline $(-8 \pm 5) \times
10^{-11}$&Lunar occultation~\cite{van75}&-1.333&1.6\\ \hline
$(-6.4 \pm 2.2) \times 10^{-11}$&Lunar tidal
acceleration~\cite{van75} &-0.8421&1.4328\\ \hline $-15.30 \times
10^{-11}$&Early Dirac theory~\cite{bla78}&11.7692 &2.0582\\ \hline
$-5.1 \times 10^{-11}$&Additive creation
theory~\cite{bla78}&-0.5730&1.2644\\ \hline $(-16 \pm 11) \times
10^{-11}$&Multiplication creation theory~\cite{fau76}
    &8.00 &2.0869\\
\hline $-4.0 \times 10^{-13}<\frac{\dot G}{G}<+3.0 \times
10^{-13}$& Big Bang Nucleosynthesis~\cite{cop04dk}&-0.0003
&-0.0021\\ \hline $-2.5 \times 10^{-10}\leq\frac{\dot
G}{G}\leq+4.0 \times 10^{-11}$&WDG 117-B15A~\cite{ben04al}&-3.0
&1.8\\ \hline$\left|\frac{\dot G}{G}\right| \leq +4.10 \times
10^{-10}$&WDG 117-B15A~\cite{gli65}&1.1319 &3.09\\ \hline$-(0.6
\pm 4.2) \times 10^{-12}$&Double-neutron star
binaries~\cite{tho96}&-0.0043 &0.0254\\ \hline$(0.46 \pm 1.0)
\times 10^{-12}$&Lunar Laser Ranging~\cite{tur04al}&0.0318
&-0.2110\\ \hline $1 \times 10^{-11 \pm 1}$&Wu and
Wang~\cite{wu86w}&0.0666 &-0.5\\ \hline

\end{tabular}
\end{minipage}
\end{table*}

It may be mentioned that, being motivated by LNH, Ray and
Mukhopadhyay~\cite{ray05m} have confronted theoretically derived
values of $\dot G/G$ obtained by solving Friedmann equations (for
flat model) with time dependent $G$ and time dependent $\Lambda$
of phenomenological character for testing the plausibility of
experimentally determined values of variation of the gravitational
constant. It has been shown by them that for certain values of the
parameters of the models, theoretically and experimentally
determined values agree with each other. Moreover, it has been
found that $G$ decreases with time as suggested by
Dirac~\cite{dir37}. Starting from Raychaudhuri's
equation~\cite{ray79} for the Brans-Dicke theory~\cite{bra61d},
Berman~\cite{ber92} has derived an expression for $\dot G/G$
involving the deceleration parameter $q$ and density parameter
$\Omega$. It has been shown there that for no variation of $G$,
one must have $2q=\Omega$. Now, for an accelerating universe $q<0$
whereas $\Omega$ is always positive. This implies that time
variation of $G$ is guaranteed by an accelerating universe.

 \section{Discussions}
Ever since the inception of the idea of large numbers, it has
drawn attention of researchers in various fields ranging from
geophysics and earth science to astrophysics and cosmology. In
this article we have reviewed all these works in a systematic way
to get the present status and future prospects. Here, main thrust
is given on cosmology, astrophysics and planetary science in terms
of LNH.

Now, success of a theory depends on its experimental verification,
applications in different fields and significant predictions. From
the previous sections it is clear that Dirac's LNH satisfies all
these criteria. LNH related works have an interdisciplinary flavor
in the sense that researchers of various fields have worked on
this particular topic in a unique manner. Not only that, some of
the works done on LNH from different standpoint seem to have a
relation between them. Berman and Trevisan~\cite{ber01ta}
investigated about a possible time variation of electric
permittivity $\epsilon_0$ and magnetic permeability $\mu_0$ and
showed that total density of the universe $\rho$ is approximately
proportional to $t^{-2}$. Ray and Mukhopadhyay~\cite{ray05m}
considered a flat Friedmann model with variable $\Lambda$ and $G$.
It is interesting to note that time variation of $\rho$
in~\cite{ray05m}, for a small value of the parameter $\alpha$,
will be similar to that [34]. Again, in another work of Berman and
Trevisan~\cite{ber01tb} regarding time variation of the
gravitational constant, velocity of light etc., it is shown that
$\rho \propto t^{-1}$ which may also be obtained from the work of
Ray and Mukhopadhyay~\cite{ray05m} for negative $\alpha$, i.e.
attractive $\Lambda$. In this regard it is to be noted here that
Gilson~\cite{gil04} has proposed a model in which necessity of
dark matter for explaining the flat galactic rotational curve is
alleviated by invoking the idea of an attractive $\Lambda$
producing a decelerating universe. It is already mentioned that
negative $\Lambda$ can be obtained from the work of Ray and
Mukhopadhyay~\cite{ray05m} for a negative $\alpha$ without
creating any disturbing physical feature relating the scale
factor, energy density and gravitational constant.
Shemi-zadeh~\cite{she02}, starting from various large number
coincidence and making some suppositions involving the fine
structure constant, the Hubble parameter, the Planck frequency
etc., has derived simple expressions for the age of the universe,
the deceleration parameter, the Hubble parameter etc. According to
Shemi-zadeh~\cite{she02}, his work can be extended further with
suitable physical models which may be connected to the Brane world
models. It may be mentioned here that unlike most of the
literature where $G \propto t^{-1}$, Milne~\cite{mil35} found that
$G$ is directly proportional to the cosmic age $t$. Recently,
using Dirac's LNH, Belinchon~\cite{bel02} has arrived at the same
relation between $G$ and $t$ as that of Milne~\cite{mil35}. In his
bulk viscous model (although not directly related to or motivated
by LNH) with time-varying $\Lambda$ and $G$, Arbab~\cite{arb03}
has shown that $G$ increases with time. In a multi-dimensional
model with an Einstein internal space and a multi-component
perfect fluid, Dehnen et al.~\cite{deh06ikm} have considered
expressions for $\dot G$. A suggestion for a possible mechanism
for small $\dot G$ in the case of two non-zero curvature without
matter is made there~\cite{deh06ikm}. For the 3-space with
negative curvature and internal space with positive curvature, an
accelerating universe with small value of $\dot G/G$ near $t_0$ is
obtained which is shown to be compatible with the exact solution
of Gavrilov et al.~\cite{gav96im}. In another example, in the same
work~\cite{deh06ikm}, with two Ricci-flat factor spaces and two
matter sources (dust + 5-brane), a sufficiently small variation of
$G$ is obtained.

 About seventy years ago, Dirac pointed out the possibility
of time-variation of a fundamental constant in the context of a
full-fledged cosmological model. It is evident from the present
review that various ideas have been germinated by LNH. Moreover,
in recent years, changeability of other constants related to
physical laws is getting more and more attention to the
researchers of various fields, viz., physics, geophysics,
astrophysics and cosmology~\cite{mar02}. Amount of variations of
some of the so called constants, viz., fine-structure
constant~\cite{web99al,web01al,pre95tm,dam96d,fuj02al,var00pi,mur01al,car01al}
and ratio of proton and electron mass~\cite{var00pi,iva02rpv} are
claimed to have been measured with reliable accuracy. It has
already been mentioned in the introduction that the erstwhile
cosmological parameter $\Lambda$, once regarded as a constant, is
now largely accepted as a function of time. These experimental and
observational results have already sparked off ideas regarding
probable variations of other fundamental constants and much more
are expected to come in near future. So, in every respect, LNH has
triggered off many new ideas, some of which may be fundamental in
nature. Recently, in a work of Shukurov~\cite{shu99} related to
interstellar matter, it has been found that the number of
Christmas dinner per galaxy turns out to be of the order of
$10^{40}$. Even the Higgs scalar-tensor theory, in which the mass
of the particles appeared through gravitational
interaction~\cite{deh91f,deh92fg,deh93f}, is also found to be
compatible with LNH~\cite{bra97}. On the other hand,
Carter~\cite{car74} is of the opinion that using anthropic
principle (both weak and strong) large number coincidence can be
explained within the framework of conventional physics and no
exotic idea, like time variation of $G$, is necessary. Thus it may
be inferred that we are still carrying a legacy of Dirac's Large
Number Hypothesis - a simple idea with profound implication.

\vspace{0.5cm}

{\bf Acknowledgment}\\ One of the authors (SR) would like to
express his gratitude to the authorities of IUCAA, Pune for
providing him the Associateship Programme under which a part of
this work was carried out. Thanks are due to A. Shukurov for
helpful discussions related to his work and also to B. K. Biswas
for going through the entire manuscript carefully.

{}
\end{document}